\documentclass[a4paper,fleqn,usenatbib]{mnras}

\usepackage[T1]{fontenc}
\usepackage{ae,aecompl}

\usepackage{graphicx}	
\usepackage{amsmath}	
\usepackage{amssymb}	
\usepackage{booktabs}

\usepackage{txfonts}

\title[ConvNets and CFHTLS lenses]{Finding strong lenses in CFHTLS using convolutional neural networks}
\author[C. Jacobs, K. Glazebrook, T. Collett, A. More, C. McCarthy]{C. Jacobs$^{1}$\thanks{E-mail:colinjacobs@swin.edu.au}, K. Glazebrook$^{1}$, T. Collett$^{2}$, A. More$^{3}$, C. McCarthy$^{4}$\\
$^{1}$Centre for Astrophysics and Supercomputing, Swinburne University of
Technology, P.O. Box 218, Hawthorn, VIC 3122, Australia\\
$^{2}$Institute of Cosmology and Gravitation, University of Portsmouth,
Burnaby Rd, Portsmouth, PO1 3FX, UK\\
$^{3}$Kavli IPMU (WPI), UTIAS, The University of Tokyo, Kashiwa, Chiba
277-8583, Japan\\
$^{4}$School of Software and Electrical Engineering, Swinburne University of
Technology, P.O. Box 218, Hawthorn, VIC 3122, Australia}

\begin{document}

\date{}
\pagerange{\pageref{firstpage}--\pageref{lastpage}} \pubyear{2016}

\maketitle

\label{firstpage}

 \begin{abstract}
We train and apply convolutional neural networks, a machine learning
technique developed to learn from and classify image data, to
Canada-France-Hawaii Telescope Legacy Survey (CFHTLS) imaging for the
identification of potential strong lensing systems. An ensemble of four
convolutional neural networks was trained on images of simulated
galaxy-galaxy lenses. The training sets consisted of a total of 62,406
simulated lenses and 64,673 non-lens negative examples generated with
two different methodologies. 
An ensemble of trained networks was applied to all of the 171 square
degrees of the CFHTLS wide field image data, identifying 18,861
candidates including 63 known and 139 other potential lens candidates. A
second search of 1.4 million early type galaxies selected from the
survey catalogue as potential deflectors, identified 2,465 candidates
including 117 previously known lens candidates, 29 confirmed
lenses/high-quality lens candidates, 266 novel probable or potential
lenses and 2097 candidates we classify as false positives. 
For the catalogue-based search we estimate a
completeness of 21-28\% with respect to detectable lenses and a purity
of 15\%, with a false-positive rate of 1 in 671 images tested. We
predict a human astronomer reviewing candidates produced by the system
would identify \textasciitilde{}20 probable lenses and 100 possible
lenses per hour in a sample selected by the robot. Convolutional neural
networks are therefore a promising tool for use in the search for lenses
in current and forthcoming surveys such as the Dark Energy Survey and
the Large Synoptic Survey Telescope.
\end{abstract}


\section{Introduction}\label{introduction}

Gravitational lensing is a consequence of the relativistic curvature of
spacetime by massive objects such as galaxies, groups and galaxy
clusters \citep{einstein_lens-like_1936, zwicky_nebulae_1937}. 
Since the discovery of the first strongly-lensed quasar by 
\citet{walsh_0957_1979}, the search for and study of strong 
lenses has become an increasingly active field of astronomy. 

Lensing phenomena can tell us much about the characteristics of distant objects
and about the universe itself across cosmic time
(\citealp{wambsganss_gravitational_1998, blandford_cosmological_1992-1}; 
see \citealt{treu_gravitational_2015} for an overview).
So-called strong lensing
occurs where the gravitational potential of a lensing body is
sufficient, and the position of a distant source is aligned such that
multiple images of the source are produced and are detectable
\citep{schneider_gravitational_2006, treu_strong_2010}.

Strong lenses have
numerous scientific uses including constraining the mass content and
density profiles of galaxies for both dark matter and baryons
\citep{treu_internal_2002-1, bradac_b1422+231:_2002, treu_massive_2004, sonnenfeld_sl2s_2015, oguri_stellar_2014, auger_sloan_2010, collett_core_2017}
and allowing us to study otherwise undetectable young galaxies at high
redshift when magnified by these gravitational telescopes
\citep[e.g.][]{newton_sloan_2011, quider_ultraviolet_2009, zheng_magnified_2012}.

Strong lenses are also valuable to cosmologists, as they allow
allowing us to constrain several cosmological parameters. This includes the
Hubble constant and dark energy equation of state, particularly when time
delay information is present such as in the case of lensed quasars
\citep{refsdal_possibility_1964, tewes_cosmograil:_2013, suyu_two_2013, oguri_sloan_2012, collett_cosmological_2014, bonvin_h0licow_2016}.

Only several hundred high-quality examples of galaxy-galaxy strong
lenses are known (\citealt{collett_population_2015}, the Masterlens
database\footnote{Database of confirmed and probable lenses from all
  sources, curated by the University of Utah.
  http://admin.masterlens.org}), but current and next-generation wide
surveys are likely to capture many more.
\citet{oguri_gravitationally_2010} predict some 8000 lensed quasars are
to be found in Large Synoptic Survey Telescope (LSST)
\citep{ivezic_lsst:_2008} imaging and of order 1000 in the Dark Energy
Survey \citep{the_des_collaboration_dark_2005} (DES)\footnote{http://www.darkenergysurvey.org}.
\citet{treu_strong_2010} predicts \textasciitilde{}1 lens per square
degree with ground-based telescopes and \citet{collett_population_2015}
performed modelling suggesting up to 2400 galaxy-galaxy
strong lenses should be identifiable in DES given an optimal stacking
strategy, and of order \(10^5\) should be detectable in the LSST and
Euclid \citep{amiaux_euclid_2012} survey databases.

Identifying strong lenses from amongst the many millions of non-lensing
galaxies in the surveys of the next decade presents an interesting
challenge. Lenses can be complex, spanning a range of morphologies,
sizes and colours. Describing them, whether in code or for a human
audience, is difficult to do without ambiguity.

With survey databases now in the terabyte-petabyte range, visual search
of each potential lens galaxy by a human astronomer is no longer a
feasible option, and so we need algorithms to find candidates for us.
Previous automation strategies have included searching for features
characteristic of strong lenses such as arcs and rings
\citep{lenzen_automatic_2004, alard_automated_2006, estrada_systematic_2007, seidel_arcfinder:_2007, more_cfhtlsstrong_2012, gavazzi_ringfinder:_2014},
fitting geometric parameters measuring an arc shape and searching for
blue residuals in galaxy-subtracted images.

\begin{figure*}
\centering
\includegraphics[width=0.95000\textwidth]{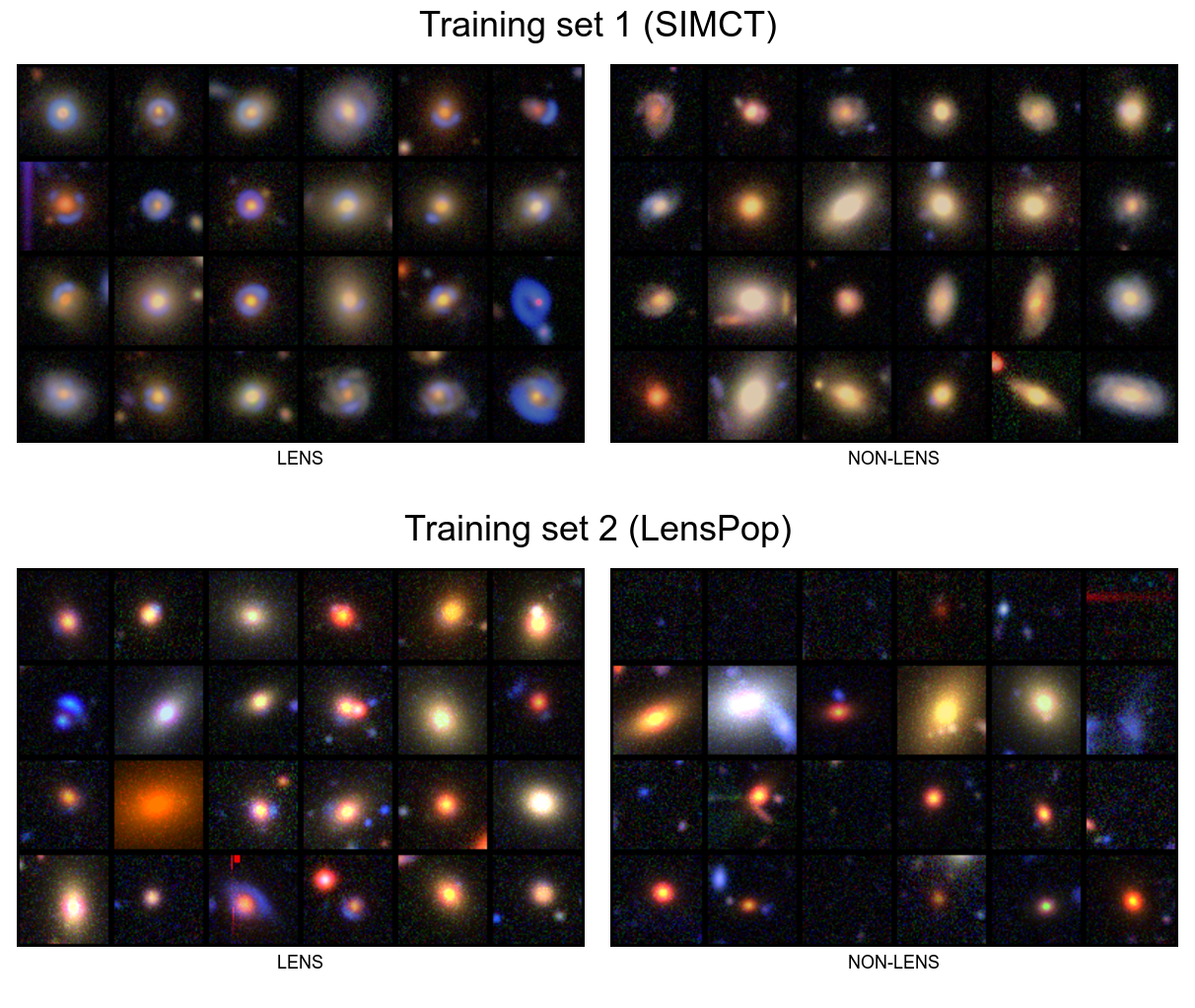}
\caption{Example synthetic lens images from two training sets. Top:
synthetic lensed sources on real survey galaxies. Bottom: Synthetic
source and deflector, on real survey backgrounds.}\label{fig:tsets}

\end{figure*}

\citet{marshall_automated_2009} and \citet{brault_extensive_2015}
modeled potential lens galaxies candidate as a strong lens system and
used the model's fit to the data as an estimate lens candidate
likelihood.

Other approaches include using principal component analysis to subtract
galaxies from imaging data \citep{joseph_pca-based_2014}; parameterising
the colour and shape of lensed quasars \citep{chan_chitah:_2015}; and
training thousands of citizen scientists on simulated images and then
having them visually inspect cutouts \citep[hereafter SWI and
II]{marshall_space_2016, more_space_2016}. 

Each methodology has successfully identified tens to hundreds of new
lenses or high-quality candidates for follow-up (e.g.\textasciitilde{}59 
in SWII). The problem faced by any automated strategy is the wide variety of
morphologies present in strong lens systems and the potential for
confusion with non-lenses, such as blue spiral galaxies. The difficulty
in paramaterising the full range of potential lenses has meant that
inspection by human experts has remained both a cornerstone and
bottleneck in lens-finding efforts.

In contrast to traditional approaches, machine learning classification
techniques forego human intuitions regarding the weighting and
relationships of significant features of the dataset and instead
algorithmically extract a useful parameterisation from patterns present
in the data only (see \citealt{jordan_machine_2015} for a brief overview
of current machine learning applications). The application of machine
learning techniques to `big data' challenges in astronomy is an active
field of research at present. Artificial Neural Networks (ANNs) in
particular have been successfully applied to astronomical problems.
\citet{dieleman_rotation-invariant_2015} used convolutional neural
networks to replicate human judgements about Sloan Digital Sky Survey
(SDSS) galaxy morphology as part of a GalaxyZoo Kaggle
challenge\footnote{https://www.kaggle.com/c/galaxy-zoo-the-galaxy-challenge}
and \citet{huertas-company_catalog_2015} applied them to morphological
classification in the CANDELS fields. \citet{hoyle_measuring_2016} used
deep learning (see Section~\ref{sec:background}) techniques for
photometric redshift estimation. \citet{schawinski_generative_2017}
employed Generative Adversarial networks
\citep{goodfellow_generative_2014-1} to develop a novel deconvolution
technique and recover features in SDSS galaxy images. In the
lens-finding arena, \citet{bom_neural_2017} employed neural networks and
the Mediatrix Filamentation Method, while \citet{ostrovski_vdes_2017}
applied Gaussian Mixture Models to the lensed quasar search.

The success of these machine learning techniques in other areas of
astronomy, and more broadly in computer vision (i.e. the identification 
and classification of objects in digital images) makes them a promising
candidate for learning and detecting the particular morphology of strong
lenses. However, inherent in all supervised learning methodologies, 
especially in convolutional neural networks, is the need for a large,
diverse and fully-labelled training set with which to tune the network's
weights. Training sets for complex image classification networks are
typically of the order of \(10^6\) images and anything smaller than
\(10^4\) images is unlikely to suffice for robust training of a complex
network which may have order \(10^7\) to \(10^9\) parameters to be
optimized. The work is further complicated by the rarity of strong
lensing systems, with the implication that any sub-optimal classifier
will overwhelm good potential lenses with false positives.

In this work we develop the application of convolutional neural networks
to detecting the morphology of galaxy-galaxy lenses in optical images.
We focus on developing a synthetic training set of
sufficient size and quality, and on ensuring that false positives were
minimised. We use the lens quality grade scale as outlined in SWII, with
images 0) unlikely to contain a lens, 1) possibly containing a lens, 2)
probably containing a lens and 3) almost certainly containing a lens. In
this paper we define false positives as any candidates that we judge to
be grade 0.

The Canada France Hawaii Telescope Legacy Survey (CFHTLS) was chosen for
our search as it has been extensively searched for strong lenses
previously, providing us with an opportunity to benchmark the
performance of this approach. As well as visual searches
\citep{elyiv_search_2013, sygnet_search_2010} and
serendipitous discoveries \citep{thanjavur_phd_2009} there have been
several previous robotic searches of the entire survey as part of the
CFHTLS Strong Lensing Legacy Survey (SL2S) \citep{cabanac_cfhtls_2007}.
\citet{more_cfhtlsstrong_2012} used the \textsc{arcfinder} robot and
\citet{gavazzi_ringfinder:_2014} used the \textsc{RingFinder} algorithm
to detect galaxy and cluster-scale strong lenses. These two searches, in
addition the \textsc{SpaceWarps} search, have collectively identified
over 500 lenses and potential lenses which we can use to assess the
effectiveness of our own algorithm. These searches and the lenses they
discovered are detailed in Section~\ref{sec:known_sources}.

In this paper we present our convolutional neural network-based lens
finder and apply it to the CFHTLS survey. The paper is organised as
follows: Section~\ref{sec:background} provides a brief overview of
artificial neural networks and the training process that powers them,
and details the results of previous automated lens searches in the
survey. In Section~\ref{sec:method} we describe the challenges of lens
finding and how we assemble our networks and the simulated lens training
sets to train them, as well as the two strategies we employ in their
application to the survey data. In Section~\ref{sec:results} we apply
our lens finder to 171 square degrees of the CFHTLS survey and describe the
lens candidates we recover, including known, confirmed lenses and novel
candidates. In Section~\ref{sec:discussion} we examine the limitations
of the two survey search approaches, the use of ensembles of neural
networks, and how best to quantify the performance of the robot.
Finally, in Section~\ref{sec:conclusions} we summarise the performance
of our lens finder and provide a brief outlook of how future work may
improve its performance and usefulness to astronomy.

\section{Background}\label{sec:background}

Artificial Neural Networks (ANNs) were first described as far back as
the 1950s \citep{rosenblatt_perceptron-perceiving_1957} and
were refined over decades, but fell into relative disuse in favour of
other algorithms due to performance and scaling issues. More recent
advances in the algorithms, including improved network initalisation
\citep{hinton_reducing_2006}, optimized gradient descent methods
\citep{duchi_adaptive_2011, zeiler_adadelta:_2012}, dropout
\citep{hinton_improving_2012}, as well as improvements in hardware and
the availability of large labelled data sets, have enabled ``deep
learning'' - training ANNs with many layers - and a resurgence in their
use. In recent years the field of computer vision has been
revolutionized by advancements in deep learning techniques, in
particular the use of convolutional neural networks 
(``ConvNets'' -  \citet{fukushima_neocognitron:_1980, lecun_gradient-based_1998}).
In 2012 \citet{krizhevsky_imagenet_2012}, employing a convolutional neural
network of novel design, entered the benchmark ImageNet Large-Scale
Visual Recognition Challenge\footnote{An annual competition considered
  the benchmark in computer vision performance, with a training set of
  \(~10^6\) images across 1000 categories.
  http://www.image-net.org/challenges/LSVRC/} \citep[ILSVRC - ][]{russakovsky_imagenet_2015}
 and were able to surpass in accuracy
the state-of-the-art of conventional computer vision techniques by 10\%,
an impressive margin. Since then, the development and application of
convolutional neural networks remains an active area of research
\citep{fleet_visualizing_2014, szegedy_scalable_2014, simonyan_very_2014, he_deep_2015}.

Artificial neural networks, named by analogy to the neuronal network in
the brain that inspired their design, are constructed as a layered
network of interconnected nodes ("neurons"), each of which takes a weighted vector
as inputs and outputs a scalar value passed through an activation function - a 
differentiable function that introduces non-linearity into the network. 
At the first layer, the raw inputs
are weighted and passed as inputs to a layer of input neurons. This is
repeated for an arbitrary number of intermediate ``hidden'' layers. At
the last layer, the outputs are interpreted as an approximation function
appropriate to the problem domain, such as the probability that the
input is member of a particular class. For a network of sufficient size
and complexity, arbitrary logic can be represented in the connections
between artificial neurons, encapsulated by the weights that
parameterise the network.

Convolutional neural networks extend ANNs by taking advantage of the
spatial relationship between input values, namely, the pixels in an
image. In a convolution layer, instead of fully connecting all neurons,
groups of input neurons are applied to small regions of neighbouring
pixels, and the associated weights are shared amongst many groups. This
approach has several desirable properties. It vastly reduces the number
of weights that characterise the network, and exploits key
characteristics of the visual domain such as the spatial significance of
neighbouring pixels and translational invariance. These small groups of
weights (kernels) are convolved with the image and act as feature
detectors. At earlier layers they perform the function of detecting
simple geometric features such as edges or patches of colour; at later
levels the hierarchy of features becomes increasingly abstract. In 
standard computer vision applications such as everyday object classification, 
a network may first activate on patterns such as curves and corners; 
then features that resemble more complex shapes such as wheels and doors; 
and finally recognise the arrangement of these features as a truck. 

Strong Lenses too have a particular and distinctive morphology, 
information which must be used by any automated lensfinder (at least 
in the absence of quality spectroscopic data). For this reason
we test the most successful computer vision technique developed
to date. We explore whether the algorithm can develop a similar feature 
hierarchy that is useful in the astronomical context of strong lens 
finding. 

In assessing the quality of a sample produced by a network, the
astronomical terminology of purity and completeness differ in their
definitions slightly from the terms used in the computer science
literature. In this work we use these conventional astronomical terms
where possible. A definition of these terms, and others used to describe
the performance of the classifier, is presented in Table~\ref{tbl:terminology}.

\hypertarget{tbl:terminology}{}
\begin{table*}
\centering

\caption{\label{tbl:terminology}Comparison of terminology in machine
learning and astronomy as applied to sets of candidate objects. }

\begin{tabular}{|p{7cm}|p{7cm}|}
\toprule

Astronomical term & Machine learning term \\\midrule

\textbf{Purity}: The fraction of the returned sample that consists of
genuine examples of the objects studied. & \textbf{Precision}: The
fraction of the examples identified by the machine learning algorithm
that are true positives. \(\frac{TP}{TP + FP}\) \\
\textbf{Completeness}: The fraction of the genuine objects studied that
are included in the returned sample. & \textbf{Recall}: The fraction of
the positive examples identified by the machine learning algorithm
(\(\equiv\) \textbf{True Positive Rate}): \(\frac{TP}{TP + FN}\) \\
\ & \textbf{False Positive Rate}: The fraction of negative examples
classified incorrectly by the machine learning algorithm:
\(\frac{FP}{FP + TN}\) \\
\ & \textbf{Accuracy}: The fraction of examples classified correctly by
the machine learning algorithm: \(\frac{TP + TN}{total size}\) \\

\bottomrule
\end{tabular}

\end{table*}

\subsection{Training Neural Networks}\label{sec:training_ann}

The essence of machine learning techniques is the algorithmic extraction
of patterns from data. Supervised learning techniques, such as that
presented here, take a labelled (i.e.~pre-classified) data set and use
it to iteratively optimize a function that parameterizes the differences
between the classes of data presented. The training of ANNs involves the
optimization of a loss function using the techniques of backpropagation
\citep{hecht-nielsen_theory_1989} and stochastic gradient descent
\citep{bottou_large-scale_2010}. In brief: mathematically we consider
our neural network as a large differentiable function that is
parameterized by a set of weights \(\mathbf{W}\) which, for any input
vector \(\mathbf{x}\), outputs a vector
\(\mathbf{y} \in \mathbb{R}^{N}\) which we interpret as the
probabilities that \(\mathbf{x}\) is a member of a defined set of \(N\)
categories: \begin{equation}
\mathbf{y} = F(\mathbf{x}, \mathbf{W}) 
\label{eq:train1}\end{equation} We then take a training set
\(\mathbf{\hat{X}}\) consisting of of \(k\) examples, together with
labels of known correct categories \(\mathbf{\hat{y}}\). We define a
loss function \(L\): \begin{equation} 
    L = L(\mathbf{y}, \mathbf{\hat{y}})  \textrm{ where } \mathbf{\hat{y}} = F(\mathbf{\hat{X}}, \mathbf{W})
\label{eq:train2}\end{equation} such that
\(L \approx 0 \ \text{if}\ \mathbf{y} = \mathbf{\hat{y}}\) (all
examples classified correctly with probability 1) and \(L\) increases as
the number of mis-classifications on the training set increases. A cross
entropy loss function per \citet{cao_learning_2007} is a conventional choice.  
The problem then
becomes to minimize the function \(L\) by finding the optimum set of
weights \(\mathbf{W}\) given our training set \(\mathbf{\hat{X}}\). We
do this using a gradient descent method, iteratively calculating the
gradients \(\frac{\partial L}{\partial W_i}\) for each \(\mathbf{x}\) in
\(\mathbf{\hat{X}}\) and \(W_i\) in \(\mathbf{W}\), updating
\(\mathbf{W}\) accordingly until we converge
to a minimum value of \(L\).
The back-propagation algorithm for calculating the weight gradients and
the stochastic gradient descent method used to minimize \(L\) and
converge on a set of weights are described fully in
\citet{lecun_backpropagation_1989}.

\subsection{Known lenses and candidates}\label{sec:known_sources}

Previous systematic searches of the CFHTLS data have yielded over 500
potential lens candidates. The CFHTLS Strong Lensing Legacy Survey
(SL2S)
\citep{cabanac_cfhtls_2007, more_cfhtlsstrong_2012, gavazzi_ringfinder:_2014, sonnenfeld_sl2s_2013}
employed several methods to identify strong lenses in the CFHTLS
imaging. \citet{more_cfhtlsstrong_2012} used the \textsc{arcfinder}
\citep{seidel_arcfinder:_2007} algorithm, visually inspecting 1000
candidates per square degree, and yielding a final sample of 127
high-quality candidates. From \citet{sonnenfeld_sl2s_2013} we include 13
further SL2S lenses confirmed with spectroscopic follow-up.
\citet{gavazzi_ringfinder:_2014}, using the \textsc{RingFinder} robot,
report a purity and completeness of 42\% and 29\% (86\% and 25\% after
visual inspection) on simulations, and present 330 good quality (grade
\textgreater{} 2) candidates from visual inspection of a sample of 2500
(13\%) returned by the algorithm on CFHTLS imaging data. In addition, we
include 55 candidates listed from earlier versions of
\textsc{RingFinder} and other searches that were not included in the
other papers. Of these SL2S candidates, 104 have follow-up indicating
definite or probable lens status.

We also include 59 new candidates identified in SWII of grades 1-3, as
discovered by 37,000 citizen scientists of the \textsc{SpaceWarps}
program. These lenses form a test set we use for optimising search
parameters as outlined below in Section~\ref{sec:test_set}.

We found 13 other confirmed lenses lying within the CFHTLS footprint had
been discovered serendipitously in Hubble Space Telescope (HST) images,
and two in SDSS spectroscopic data. These lenses are not detectable in
CFHTLS images and so we do not include them in the analysis.

Table~\ref{tbl:known_lenses} depicts the catalogue we assembled after
removing duplicates and several lenses for which we did not have imaging
available.

\subsection{Assembling a test set}\label{sec:test_set}

We use the 59 lens candidates discovered by the \textsc{SpaceWarps} citizen scientists
as a test set to aid in evaluating and optimising the performance of our robot.  
The test set consists of candidates of a high quality, which were not used in
training the robot, nor are they counted in determining the final performance metric of
confirmed CFHTLS lenses recovered. Besides the size and quality of this set,
we chose to use the SWII lenses as a
test set since, firstly, the lenses were of a size and morphology
that made finding by automated means challenging (they were not
discovered in any previous CFHTLS lens search); and secondly, they allow
for a comparison, in terms of efficiency of search time, with the
person-hours invested in training and searching by the
\textsc{SpaceWarps} volunteers who discovered them. 

\hypertarget{tbl:known_lenses}{}
\begin{table}
\centering

\caption{\label{tbl:known_lenses}The lenses and potential lenses within
the CFHTLS survey reported in previous searches which we use as a
benchmark to measure lensfinder performance. Those with reported
spectroscopic or imaging follow-up (as compiled by the Masterlens
database) we consider confirmed. }

\begin{tabular}{@{}lll@{}}
\toprule

Source & Candidates & Confirmed \\\midrule

SL2S More+12 & 117 & 57 \\
SL2S Sonnenfeld+13 & 13 & 13 \\
SL2S Gavazzi+14 & 376 & 34 \\
More+16 & 59 & 0 \\
----- & ---- & ---- \\
Total & 565 & 104 \\

\bottomrule
\end{tabular}

\end{table}

\section{Method}\label{sec:method}

Constructing our lens-finding robot and using it to identify candidates
in the survey involves the following steps. Firstly, we construct
several training sets of simulated lenses and non-lenses. We also
assemble a test set of real survey images containing the 59 SWII
lenses/candidates. We use the simulated images to train two
convolutional neural networks of our own design using the Caffe open source
deep learning framework \citep{jia_caffe:_2014}, and evaluate 
their performance on the test set.
Counting the number of lenses recovered and the false positives
identified in the test set allows us to optimise several parameters for
use in a wider search of the entire survey image dataset (``all-survey
search''). We then evaluate \(6.4 \times 10^8\) postage stamps from the
CFHTLS survey with each of the two trained networks to produce a
candidate set for grading by visual inspection. Subsequently, we adjust
the training set and network architecture and train two further neural
networks, and use an ensemble of all four of our networks in a more
restricted search of images of potential lensing galaxies 
chosen using a colour-magnitude cut from the survey catalogue. The
networks, training sets and architectures used are summarised in
Table~\ref{tbl:convnets}.

\hypertarget{tbl:convnets}{}
\begin{table}
\centering

\caption{\label{tbl:convnets}Details of the four Convnets applied to the
search. ConvNet1 and ConvNet2 were trained for the all-survey search,
then two networks with an extra convolutional layer and training sets
tailored for the catalogue-based search were trained. All four networks
were applied to the catalogue-based search. }

\begin{tabular}{@{}llll@{}}
\toprule

Network & Training set & Architecture & Ensemble \\\midrule

ConvNet1 & TS 1 & ARCH1 & All-survey, catalogue \\
ConvNet2 & TS 2 & ARCH1 & All-survey, catalogue \\
ConvNet3 & TS 3 & ARCH2 & Catalogue \\
ConvNet4 & TS 4 & ARCH2 & Catalogue \\

\bottomrule
\end{tabular}

\end{table}

\subsection{Training Data}\label{sec:training}

In order to apply convolutional neural networks to the lens-finding
problem, we must first assemble a training set to optimise the network.
The number of known galaxy-galaxy strong lenses is of order \(10^2\),
which practice suggests is several orders of magnitude below the minimum
required to usefully train a ConvNet. Furthermore, images of known
lenses stem from a heterogeneous mix of surveys, instruments and bands.
We must employ an alternate strategy to assemble a training set of
appropriate size and consistency. Fortunately, gravitational lensing can
be readily modelled using basic physical principles. Here we take the
approach of generating synthetic data - simulated lenses in sufficient
quantity to form a practical training set.

We assume that, if the artificial lenses are realistic enough, then
the network will learn lensing features, such as colour and geometry,
sufficiently well to generalize successfully to real astronomical data.
Conversely, any errors or biases in the training set are likely to be
reflected in the ConvNet's performance, not necessarily in predictable
ways. For lens finding, we require that the lensing galaxies and sources
should match the colour, size and shape distributions of the real
universe as closely as possible; the training set should be adapted for
the seeing and filters of the particular target survey, in this case
CFHTLS; noise (shot and sky) should match the realistic values for the
integration times present in the survey imaging.

Free parameters include the number of examples, the format and
dimensions of each training image, the location of lenses in the images
and the composition of the negative examples (i.e.~non-lens training
images).

We employ two strategies to assemble training images conforming to the
above constraints. The first strategy involves identifying potential
deflector galaxies in the survey, and adding a simulated lensed source
galaxy to the survey image. This method is implemented in the
\textsc{SIMCT}\footnote{https://github.com/anupreeta27/SIMCT} pipeline
described in detail in SWII but can be summarized
as follows. Potential deflector galaxies, primarily large, massive
early-type galaxies (ETGs), are selected from the CFHTLS catalogue by
brightness, colour and photometric redshift. A model mass distribution
is assigned to each galaxy using a singular isothermal ellipsoid profile
and aligned with the image. Artificial sources are constructed according
to the observed parameters of redshift, luminosity and size and given
plausible surface brightness profiles and ellipticities. The multiple
images of the source are then generated and added to the deflector image
using GRAVLENS \citep{keeton_computational_2001} raytracing code with
noise and seeing simulated to match those present in the CFHTLS. As the
number of synthetic lens examples is limited to the lens candidate
catalogue, the data is augmented by applying three 90 degree rotations and
a random translation of \(\pm\) 10 pixels in each axis to create four
training images per catalogue galaxy.

Our second strategy for generating a training set is to generate mock
images where both the lens and source are simulated. This has the
advantage that the number of examples that can be generated is
effectively infinite, but has the disadvantage that noise, seeing and
artifacts must also be simulated to well match real survey data. We do
this using a modified version of the \textsc{LensPop}\footnote{https://github.com/tcollett/LensPop} code
\citep{collett_population_2015}. \textsc{LensPop} uses the observed
velocity dispersion function of elliptical galaxies
\citep{choi_internal_2007} to generate a population of singular
isothermal ellipsoid (SIE) lenses, with realistic mass, redshift and
ellipticity distributions. Lens light is then added using the
fundamental plane relation \citep{hyde_luminosity_2009} assuming a de
Vaucolours Profile and the spectral energy distribution of an old,
passive galaxy. Sources are assumed to have an elliptical exponential
profile, with redshifts, sizes and colours drawn from the simulated
faint source catalogue of \citet{connolly_simulating_2010}.

In principle \textsc{LensPop} can simulate lenses with extremely faint
arcs or extremely small Einstein Radii. In practice such systems are
undetectable as lenses and will not contain sufficiently strong features
to usefully train the ConvNets. We therefore adopt the detectability
criteria defined in \citep{collett_population_2015}, with a signal-to-noise 
ratio >20, magnification of 3 or greater and a resolution threshold calibrated
such that arcs and counter-images can be resolved from each other and the lens.
Only these detectable lenses are used to form our training set.

We modified \textsc{LensPop} to generate mock images with appropriate
seeing (0.8" in all bands) and shot noise for the CFHTLS, but without
readnoise or shot noise from the sky background. These mock images were
then superimposed on backgrounds chosen randomly from a tile of the
CFHTLS survey. Non lenses were generated in the same way but with the
source fluxes set to zero; this results in images with synthetic
early-type galaxies drawn from the same distribution as our lenses but
with no source light.

The redshift distributions of the simulated lenses and sources are as
per \citet{collett_population_2015} and SWII.

For the whole-survey search, where every pixel of the image database was
tested, the ConvNets are required to distinguish strong lensing systems
from any other object visible in the survey sky. In the case of the
first training set, negative examples consist of images centered on an
assortment of galaxies (ellipticals, spirals and irregulars) with no
lensed source (Figure~\ref{fig:tsets}). For the second training set,
we test a different strategy, with negative examples containing 
empty sky, spiral galaxies, stars, and artifacts, drawn from a 
random position on a survey tile as 60x60 pixel
stamps from the survey imaging.\footnote{Since the negative examples
were chosen at random from the survey images there is a chance that a 
few lens candidates could be included. Since their effect on the network
is in proportion to their number in the training set, this is unlikely to
be a significant problem.} The two training sets (TS1 and
TS2) are outlined in Table~\ref{tbl:testsets}. 20\% of the images in
each training set are excluded from training and set aside as validation
sets for measuring accuracy during the training process.

\hypertarget{tbl:testsets}{}
\begin{table}
\centering

\caption{\label{tbl:testsets}Summary of the image sets used to train the
corresponding convolutional neural networks. SIMCT simulations use real
survey ETGs and simulated sources; LensPop simulates both source and
deflector. }

\begin{tabular}{@{}lllll@{}}
\toprule

Training set & Simulations & Pos examples & Neg
examples & Total \\\midrule

TS1 & \textsc{SIMCT} & 6657 & 7813 & 14470 \\
TS2 & \textsc{LensPop} & 11799 & 12910 & 24709 \\
TS3 & \textsc{SIMCT} & 3950 & 3950 & 7900 \\
TS4 & \textsc{LensPop} & 40000 & 40000 & 80000 \\

\bottomrule
\end{tabular}

\end{table}

For the catalogue-based search, all tested images are postage stamps of
pre-selected red elliptical galaxies and so the training sets are
constructed accordingly. Two further training sets are constructed. The
first (TS3) uses \textsc{SIMCT} simulated lenses with a small random
translation of 0-10 pixels. The negative training set consists of an
equal number of ETGs randomly selected from our catalogue. The second
(TS4) consists of 40,000 \textsc{LensPop}-simulated lenses, again with
up to 10-pixel scatter from the centre of the image, and an equal number
of catalogue ETGs as the negative training set (unlike TS2, which included
stars, empty sky, etc. in the negative set). The spatial jitter, added
to TS3 and TS4, is introduced in order to prevent the ConvNets from
developing an over-sensitivity to position in the image.

Determining the optimal size of a training set is not straight forward,
as it depends on the number of weights to be trained and the complexity
of the features the network must learn. Large-scale visual recognition
applications typically require training sets in the millions. We assume
that, given the small number of convolutional layers (as outlined below)
and classes (two) that a training set of order a few times \(10^4\) to
\(10^5\) examples will suffice to train our networks.

\subsection{Network Architecture and
Pipeline}\label{network-architecture-and-pipeline}

\begin{figure}
\centering
\includegraphics[width=\columnwidth]{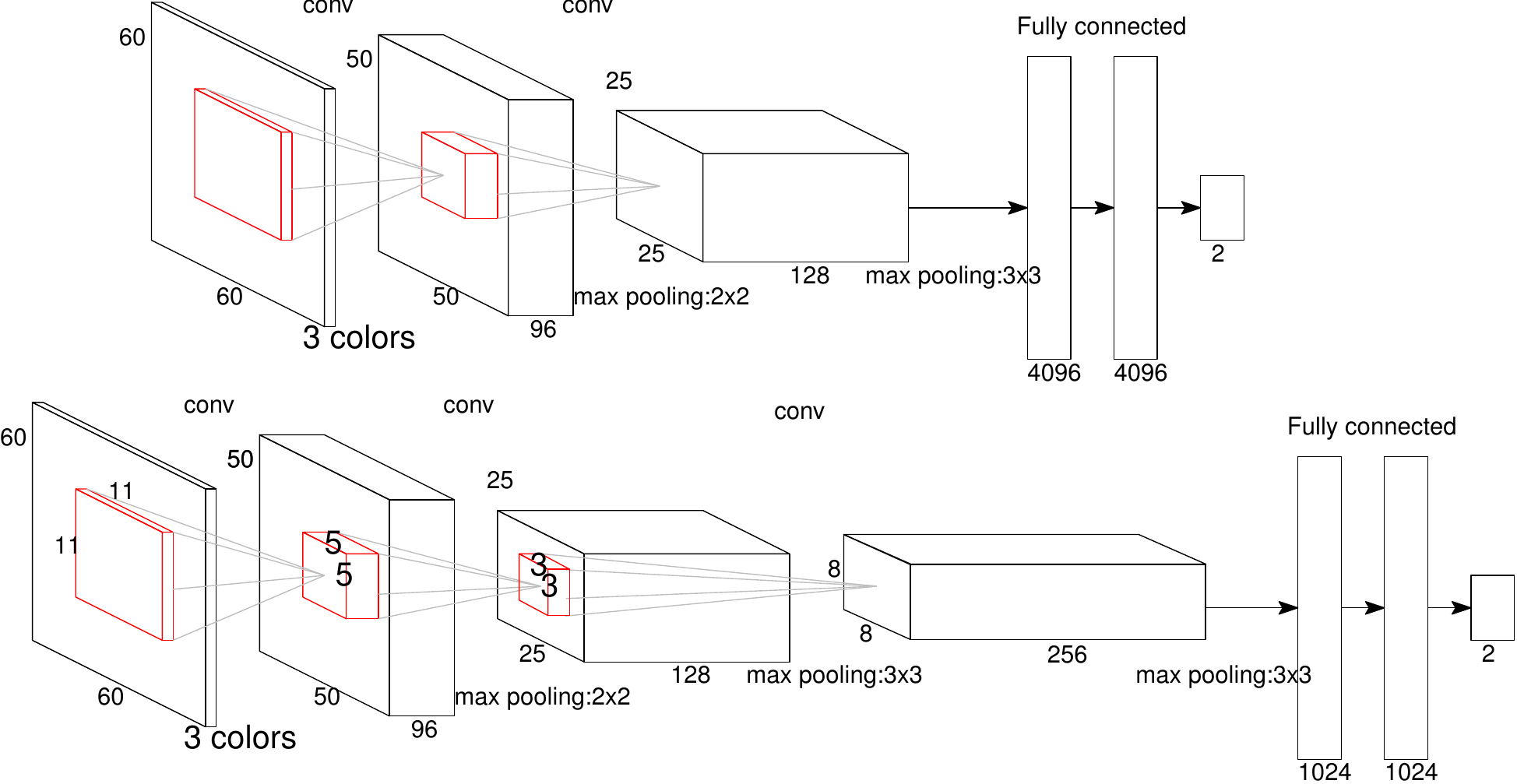}
\caption{Convolutional network architectures. The networks consist of a)
two convolutional layers with kernel sizes 11x11 and 5x5, with 96 and
128 feature maps respectively (ARCH1); and b) three convolutional layers
with kernel sizes 11x11, 5x5 and 3x3 with 96, 128 and 256 feature maps
(ARCH2). Between each layer is a ReLu activation layer and dropout of
0.5 is applied before each fully connected layer.}\label{fig:network}

\end{figure}

The convolutional neural networks were trained using the Berkeley Caffe
deep learning package. Two network architectures
are used, the first (ARCH1) consisting of two convolutional layers and
two fully connected layers of 4096 neurons each, the second (ARCH2)
consisting of three convolutional layers and two fully-connected layers
of 1024 neurons each (Figure~\ref{fig:network}). These networks with two
to three convolutional layers can be compared to the five of the ILSVRC
Alexnet \citep{krizhevsky_imagenet_2012} and up to hundreds in the more
recent literature \citep[e.g.][]{simonyan_very_2014}. A smaller network requires
less time to train, and our architecture is justified by the smaller and
morphologically simpler dataset as well as the low number of categories
- two - compared to more general computer vision applications (c.f.
ILSVRC, \(10^6\) images, 1000 categories). ARCH2 added a convolutional
layer and reduced the number of fully connected neurons; this design
resulted in a smaller number of weights to train (\(1.1 \times 10^6\),
versus \(1.8 \times 10^7\) in ARCH1). Between each layer we use a 
rectified linear unit ("ReLu") activation function \citep{nair_rectified_2010},
\(y = \text{max}(0, x)\).

The output of each
ConvNet was a softmax layer\footnote{A normalised exponential function
  that squashes an array of \(N\) real values to a \(N\)-dimensional
  vector \(\mathbf{v}\) such that \(0 < v_i < 1\) and \(\sum(v_i) = 1\):
  \(\sigma(z)_j = \frac{e^{z_j}}{\sum_{n=1}^{N}e^{z_n}}\) for
  \(j = 1, ..., n\).} resulting in a real-valued number
\(0 \le s \le 1\) which we interpret as the network's confidence that
the candidate image is a lens.

The networks are trained with input images of dimension 60x60 pixels in
three colours. These dimensions are chosen as, in the case of CFHTLS
imaging at .186" per pixel, this is large enough to contain the test set
lenses and small enough to prove highly performant in training and
searching. The FITS images in three bands are converted to RGB images scaled
with an arcsinh stretch using \textsc{HUMVI}
\citep{marshall_humvi:_2015} and supplied to the ConvNet as vectors of
10800 (60x60x3) floating-point numbers. Due to the optimal depth and quality 
of CFHTLS \(irg\)-band images 
we choose these bands for all images generated
for this study, consistent with SWII images seen by volunteers.\footnote{See Terapix T0007 
release explanatory document: http://terapix.iap.fr/cplt/T0007/doc/T0007-doc.pdf}

Training, when performed on a
single NVidia K80 GPU, was typically of order a few hours for ten
epochs (iterations through the entire entire training set). 
Training was conducted using stochastic gradient descent, with
Nesterov momentum of 0.9 and a decaying learning rate initialised at 0.1.
The network's weights were initialized according to the Xavier method
\citep{glorot_understanding_2010}.

\subsection{All-survey search}\label{all-survey-search}

Two networks of the same architecture (ARCH1) are trained separately on
the first two training sets outlined in Section~\ref{sec:training} and
Table~\ref{tbl:convnets}. As discussed below, an ensemble of the two is
used in identifying final candidate lenses. Our processing pipeline
constructs the set \(C\) of candidate lenses such that for any image
\(c\) with ConvNet scores \(s_{1}\) and \(s_{2}\):
\begin{equation} c \in C \ \text{if}\ s_{1}(c) > t_1 \ \text{and}\  s_{2}(c) > t_2 \label{eq:hyper}\end{equation}
where thresholds \(t_1\) and \(t_2\) are parameters chosen empirically
to achieve an optimal balance between purity and completeness. As
described below, the results presented here assume \(t_1\) = \(t_2\) =
0.95.

When processing each pointing image of 19354 pixels squared, our
pipeline divides the image into overlapping 60x60 cutouts (in three
colours) advanced with a 10-pixel stride, so that any given point in the
field will be tested at different positions in 36 cutouts. This means
that for each square degree of sky in the survey, 3.7 million
overlapping images are tested with each of the two trained ConvNets.

We filter the candidates that meet this criterion further by removing
those that are not robust under small translations. We find that
rejecting candidates that occurred in isolation, with no neighbouring
candidates in overlapping images, removed 85\% of candidates and
reduced the false positive rate accordingly.

\subsection{Catalogue-based search}\label{catalogue-based-search}

\begin{figure}
\centering
\includegraphics[width=\columnwidth]{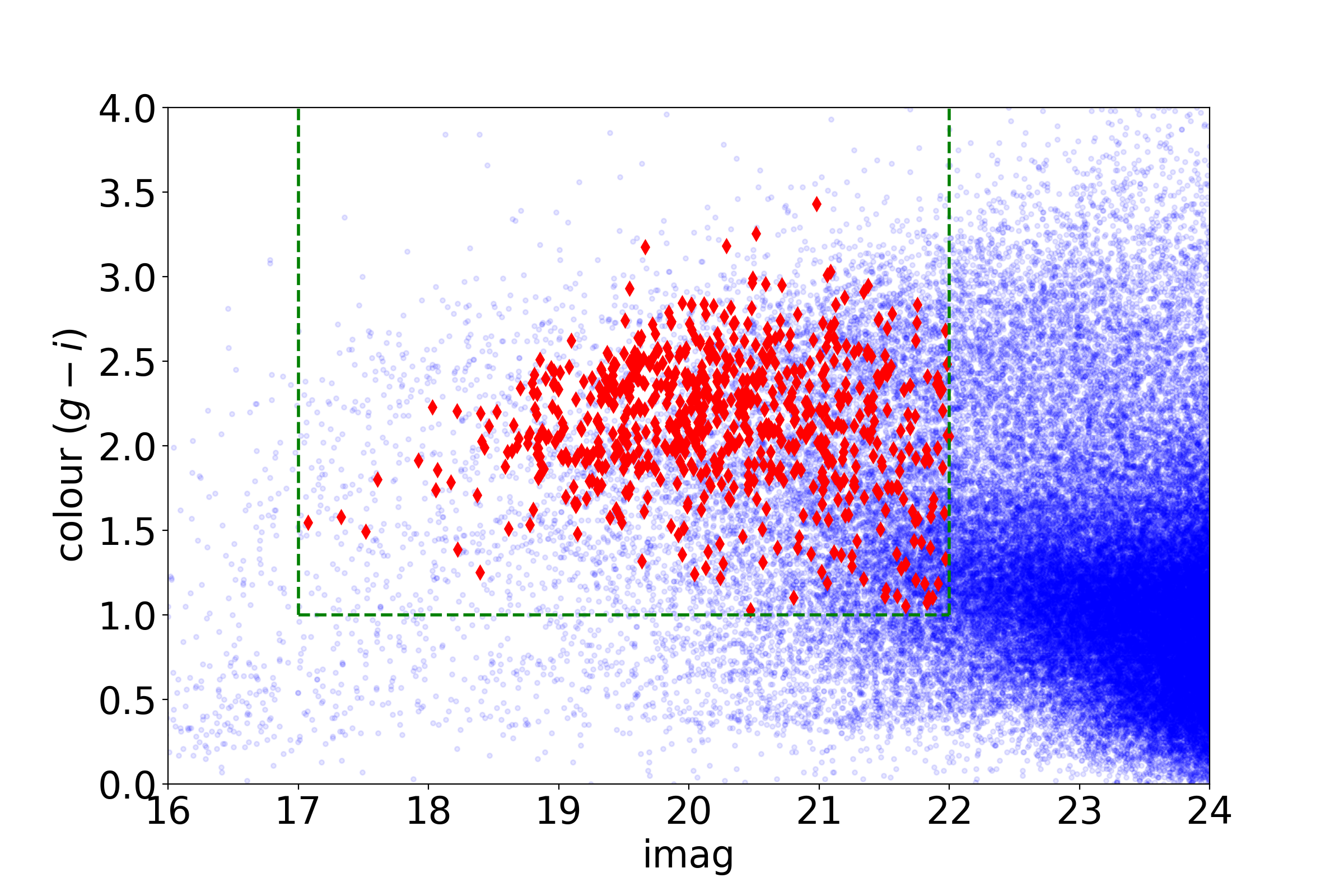}
\caption{Colours and magnitude of sources from the CFHTLS photometric
catalogue (blue dots) and sources corresponding to lensing systems and
promising candidates reported in the literature (red diamonds). Our search
considers only sources within the box marked in the green dashed line, where
(\(17 < \textrm{mag}_i < 22\)) and (\(g - i > 1\)) representing 1.4
million of the 36 million catalogued sources.}\label{fig:colormag}

\end{figure}

For the catalogue-based search, we restrict the search to catalogued
sources in a subset of colour-magnitude space consistent with likely
lensing galaxies. We expect massive ETGs to be the best candidate
deflectors due to their higher lensing potential, and so select sources
that are both bright (\(17 < \textrm{mag}_i < 22\)) and red
(\(g - i > 1\)). These thresholds are confirmed by examining the
position in colour-magnitude space of known lensing systems and
candidates within the CFHTLS footprint (see Figure~\ref{fig:colormag})
according to the 3" aperture photometry available in the survey catalogue.
98\% of these lenses are within the cut, which contains 5.6\% of the 36
million catalogue sources. We apply these cuts to the CFHTLS catalogue,
excluding those sources flagged as stars, to build our catalogue of
1,402,222 sources.

For each source in the catalogue, 60x60 pixel postage stamps centered on
the source are created and scored by each of the four ConvNets.

\section{Results}\label{sec:results}

\subsection{Simulated Lenses}\label{simulated-lenses}

\begin{figure*}
\centering
\includegraphics[width=0.90000\textwidth]{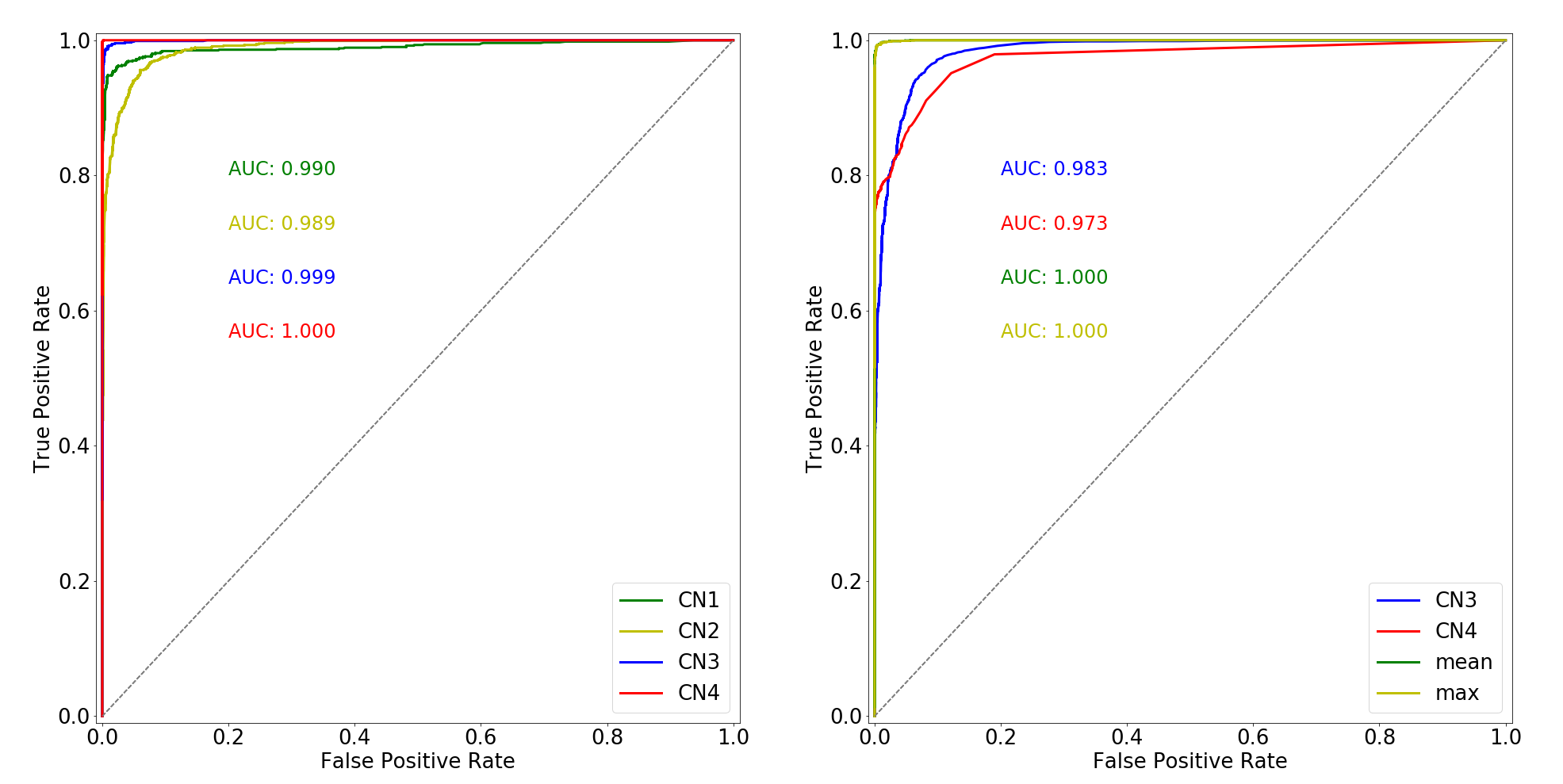}
\caption{Receiver operating characteristic curves for the trained
ConvNets, depicting the trade-off between the true positive rate
(recall) on the y axis, and the false positive rate (false positives/all
negatives) on the x axis. Each point on the curve represents a threshold
setting for the network - a lower threshold identifies more candidates
(higher recall) but also more false positives (lower
precision).}\label{fig:ROC_training}

\end{figure*}

All of the networks trained were able to distinguish simulated lenses
from non-lenses with high accuracy. On test sets not used for training,
containing equal numbers of simulated lenses and non-lenses, the four
networks trained were able achieve accuracy of 98.4\%, 91.6\%, 99.2\%,
and 99.8\%, respectively. This performance is depicted in
Figure~\ref{fig:ROC_training}. On test sets of 2000 simulated lenses,
the four networks achieved both high purity (94\%, 94\%, 100\%, 100\%)
and completeness (96\%, 95\%, 99\%, 100\%) with a score threshold of
0.5. 

The networks' performance dropped when applied to simulated lenses
generated by the method not used for training, indicating a preference
for the peculiarities of each particular simulation method.
Figure~\ref{fig:ROC_training} depicts the degraded performance for
networks 3 and 4 on each other's test sets. However, both the mean and
maximum scores of the two networks were able to classify the combined
test set with close to 100\% accuracy. This fact informed the use of a
combination of the two types of networks on the search of real survey
images as outlined below.

\subsection{All-survey search}\label{all-survey-search-1}

Two ConvNets (ConvNet1 and ConvNet2) were trained on the training sets
(described above in Section~\ref{sec:training}) and applied to a test
set consisting of 59 512x512 pixel images centered on the lens
candidates from SWII. The area covered by the test set is 
\textasciitilde{}3900 pixels squared, equivalent to 0.041 square degrees
of sky, which results in a total of 153,459 overlapping 60x60 cutouts to be tested. 
The parameters \(t_1\) and \(t_2\) from
Equation~\ref{eq:hyper} were adjusted in order to explore the balance
between true and false positives output by the system. At
\(t_1 = 0.95\), \(t_2 = 0.95\), after testing 153,459 images, the system
recovered 25/59 of the test set's lenses and returned three false
positives. These settings were used as a basis for the wider search. A subset of 
candidates returned from the test set, as well as those not detected,
are depicted in Figure~\ref{fig:foundmissed}. Some of the lenses
missed include cluster-scale lenses, red sources and possible edge-on 
spiral deflectors, which were not simulated in the training set; others
are more similar to the simulated lenses.

We applied the same pipeline to 171 square degrees of the CFHTLS survey
for which \emph{i}, \emph{r} and \emph{g} band data
were available. The results of the search with these parameters are
summarised in Table~\ref{tbl:results_main}. Classifying the images took
approximately 100 GPU-days on NVidia K20 GPUs with 2496 cores each.
After examining all of the 60x60 postage stamps with both ConvNets,
18,861 candidates were identified by the system, representing 0.1\% of
the survey area or about one source in 1900. Out of the 59 sources
in our test set of real lens candidates, the final
candidate set included 22 (37\%) of them (three of the lenses found in 
the test set search were dropped due to minor differences in position 
in the cutout between the two searches). The candidate set included a
total of 63 of 565 previously reported lenses or candidates (11\%), and
14 of the 103 confirmed lenses (14\%). Of the remaining 18,839
candidates, the authors identified 18,400 as low quality (false
positives) and 149 as potential lens candidates with grade
\textgreater{} 0. Out of 640 million images examined, this represents a
false positive rate of about 1 in 35,000. The 199 true positives
represent a purity of only 1\% for the estimated completeness of
11-14\%.

The sample of novel candidates includes six which we rate as grade
\(\ge 2\). These candidates are included in
Figure~\ref{fig:new_candidates} and Table~\ref{tbl:new_lenses}.

\hypertarget{tbl:results_main}{}
\begin{table*}
\centering

\caption{\label{tbl:results_main}Results of all-survey and catalogue searches. }
\centering

\begin{tabular}{cc} 
  \begin{tabular}[t]{@{}ll@{}}
\toprule

All-survey search\ & \ \\\midrule

Images tested & \(6.4 \times 10^8\) \\
Detection by either ConvNet & \(2.0 \times 10^7\) \\
Detection by both ConvNets & \(1.6 \times 10^5\) \\
Candidates robust under translation & \(3.5 \times 10^4\) \\
Fraction of images returned as candidates & \(2.9 \times 10^{-5}\) \\
Final candidate set & 18,861 \\
Test set lenses/candidates found & 22/59 (37\%) \\
Known lenses/candidates found & 63/565 (11\%) \\
Confirmed lenses found & 14/104 (13\%) \\
New candidates & 199 \\
False positives & 18,400 \\
Purity & 1\% \\
Completeness (known lenses) & 11-13\% \\

\bottomrule
\end{tabular}

\centering
\begin{tabular}[t]{@{}ll@{}}
\toprule

Catalogue-based search\ & \ \\\midrule

Images tested & \(1.4 \times 10^6\) \\
Final candidate set & 2,465 \\
Fraction of images returned as candidates & \(1.8 \times 10^{-3}\) \\
Test set lenses/candidates found & 23/57 (40\%) \\
Known lenses/candidates found & 117/565 (21\%) \\
Confirmed lenses found & 29/104 (28\%) \\
New lens candidates grade \textgreater{} 0 & 266 \\
New lens candidates grade \textgreater{}= 2 & 13 \\
False positives & 2097 \\
Purity & 15\% \\
Completeness (known lenses) & 21-28\% \\

\bottomrule
\end{tabular}
\end{tabular}

\end{table*}

\hypertarget{tbl:new_lenses}{}
\begin{table*}
\centering

\caption{\label{tbl:new_lenses}New potential strong lenses identified in
the CFHTLS wide fields by the ConvNet lensfinding robot. The 16 sources
consist of the candidates identified by the algorithm that have a
quality flag \(\ge 2\) and are not identified elsewhere in the
literature. The i-band magnitude supplied is from CFHTLS Terapix T0007
photometric catlog 3" aperture photmetry. Photometric redshifts for the
deflectors are provided where available. }

\begin{tabular}{@{}llllllll@{}}
\toprule

Candidate & search & source & RA & dec & grade & \(i_{mag}\) & \(z_{phot}\) \\\midrule

CN1 & all-survey & 1135\_114442 & 32.014397 & -6.962067 & 2.0 & 19.857 & 0.42 \\
CN2 & all-survey & 1134\_102957 & 32.773519 & -7.136335 & 3.0 & 20.450 & 0.83 \\
CN3 & catalogue & 1114\_035558 & 34.866662 & -5.488730 & 2.5 & 19.18 & -- \\
CN4 & all-survey & 1157\_073435 & 36.528578 & -9.983405 & 3.0 & 20.323 & -- \\
CN5 & catalogue & 1119\_077854 & 38.123480 & -6.227281 & 2.0 & 20.67 & -- \\
CN6 & both & 1120\_144248 & 37.142005 & -5.996789 & 2.0 & 20.68 & 0.84 \\
CN7 & catalogue & 1124\_121677 & 33.544766 & -6.093784 & 2.0 & 21.29 & 0.79 \\
CN8 & catalogue & 1133\_208513 & 33.367565 & -6.681572 & 3.0 & 20.53 & 0.55 \\
CN9 & both & 1156\_165804 & 37.090818 & -9.645684 & 2.0 & 19.74 & 0.30 \\
CN10 & catalogue & 1163\_117574 & 31.012133 & -9.745038 & 2.0 & 21.30 & 0.87 \\
CN11 & catalogue & 1203\_218233 & 134.733282 & -1.035681 & 2.0 & 19.48 & -- \\
CN12 & catalogue & 1309\_158230 & 217.484603 & 56.535095 & 2.0 & 20.27 & 0.55 \\
CN13 & catalogue & 1314\_020589 & 210.346656 & 55.951494 & 2.0 & 19.59 & -- \\
CN14 & both & 1328\_213514 & 210.206937 & 54.738516 & 2.0 & 20.47 & 0.59 \\
CN15 & catalogue & 1403\_048428 & 330.851737 & 3.811333 & 2.0 & 21.87 & -- \\
CN16 & catalogue & 1424\_168469 & 333.094995 & -0.303149 & 2.0 & 21.71 & 1.13 \\

\bottomrule
\end{tabular}

\end{table*}

\subsection{Catalogue-based search}\label{sec:catalog_based}

To assess the performance of the catalogue-based search, we assign each
known lens/candidate to the closest catalogue sources. Of the
\textsc{SpaceWarps} test set lenses, two were too blue and were
excluded by the colour cut leaving a test set of 57 lenses.

\begin{figure*}
\centering
\includegraphics[width=0.90000\textwidth]{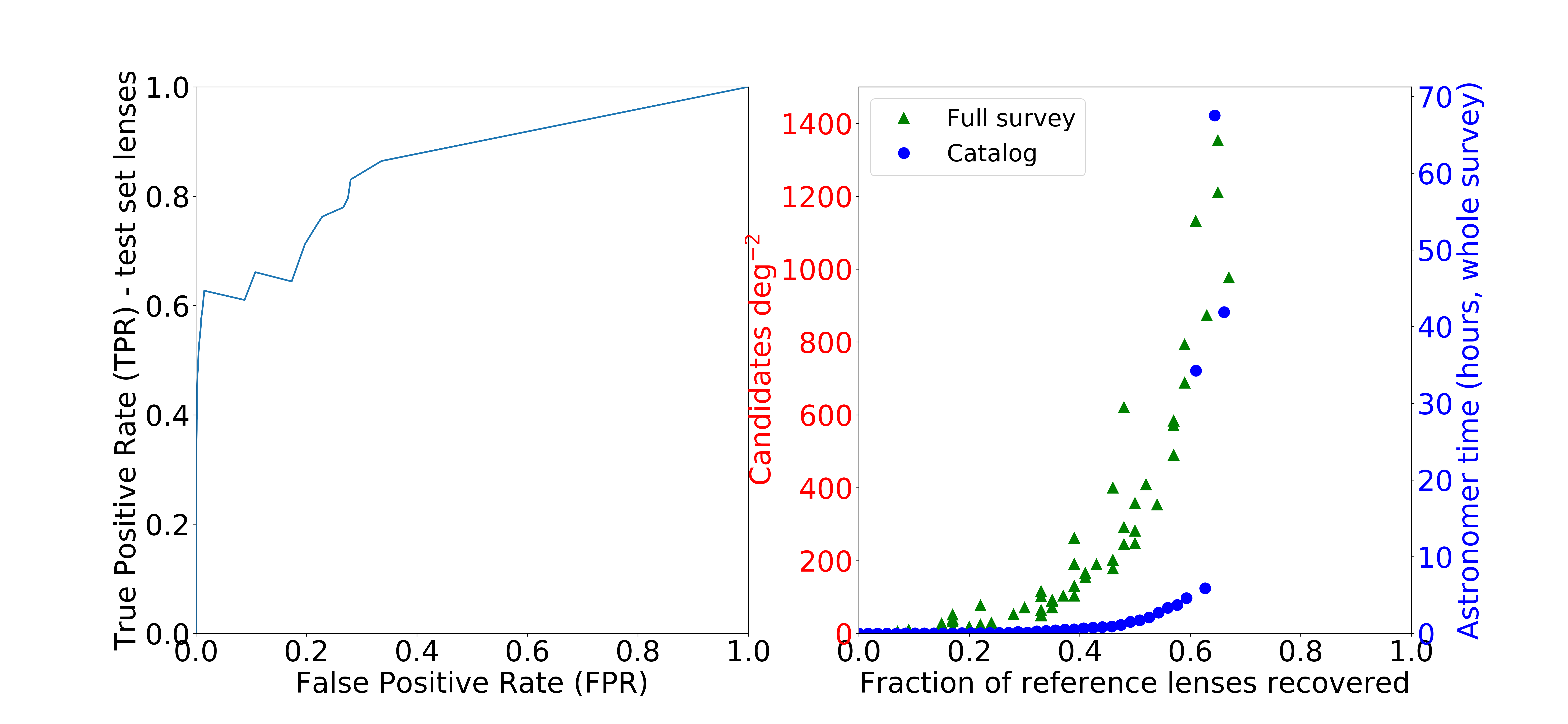}
\caption{Left: Catalogue-based lensfinder true-positive/false positive
response (ROC curve). 20\% of the test set can be recovered with a
false positive rate of .01\%; 40\% with a false positive rate of 0.17\%;
and 60\% with a false positive rate of 1.17\%. Right: Lensfinder recall
vs.~time required to inspect the produced candidate set for the
whole-survey approach (green triangles) and catalogue-based search 
(blue circles). For the whole survey, 171 square degrees, we estimate 
40\% of findable lenses
can be recovered with about 15 candidates per square degree to inspect -
or one hour of inspection time for the survey.}\label{fig:response}

\end{figure*}

For the catalogue-based search, each of the 1.4 million selected sources
was evaluated with the ConvNets 1-4. As with the whole-survey search,
the score thresholds for each ConvNet can be varied in order to assemble
subsets of the tested sources to serve as candidate sets. Each subset
can be assigned a true positive rate based on the number of test set
lenses recovered, and a false positive rate based on the size of the set
and the number of known lenses within it. The distribution of these sets
in true positive-false positive space, known in machine learning
literature as a receiver operating characteristic curve \citep[ROC -][]{fawcett_roc_2004},
succinctly describes the trade-off between precision and recall inherent
in an algorithm, and is presented in Figure~\ref{fig:response}. 

At one extreme, selected for high purity and using the four threshold 
values of (0, 0.95, 0.95, 0.15) for 
ConvNets 1-4, a candidate set of only 71 candidates was produced containing 4
of the test set lenses and 10 other known lenses or candidates. In other 
parts of the space, with low thresholds for all of the networks, candidate
sets containing hundreds of thousands of images can be produced. 
In between these two extremes, with thresholds (0, 0.95, 0.55, 0), a candidate
set of 2,465 sources is produced containing 23 of the test set lenses.
With a completeness of 40\% with respect to the \textsc{SpaceWarps}
lenses, and a candidate set small enough to inspect in under an hour,
this set was chosen as representative of likely practical settings and
was examined by eye to evaluate overall candidate quality.

\begin{figure*}
\centering
\includegraphics[width=0.90000\textwidth]{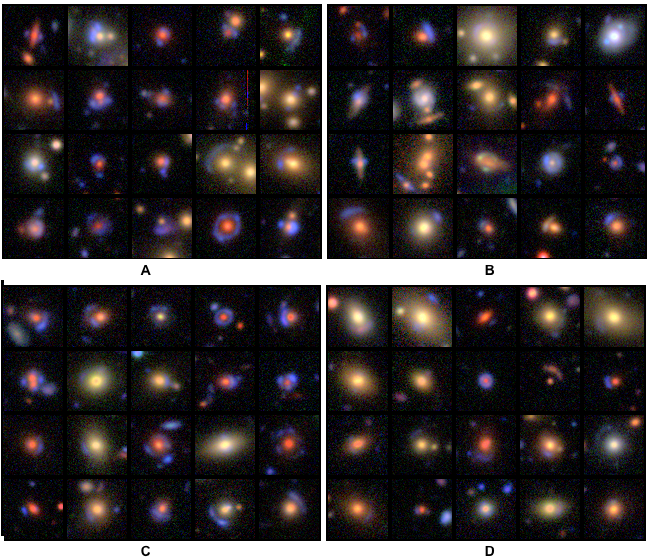}
\caption{A subset of lenses recovered by the lens finder, and those missed.
Top left, A: 20 Lens candidates identified by citizen scientists of the
\textsc{SpaceWarps} project \citep{more_space_2016} recovered by the
ConvNet robot in a candidate set of 2465 candidates. Top right, B: 20
\textsc{SpaceWarps} lenses not recovered. Bottom left, C: 20 confirmed
lenses recovered in the catalogue-based search. Bottom right, D: 20
confirmed lenses not recovered. The missed SW lenses contain more
cluster-scale lenses, redder sources and potential spiral deflectors;
the missed confirmed lenses have fainter
sources.}\label{fig:foundmissed}

\end{figure*}

\begin{figure*}
\centering
\includegraphics[width=0.90000\textwidth]{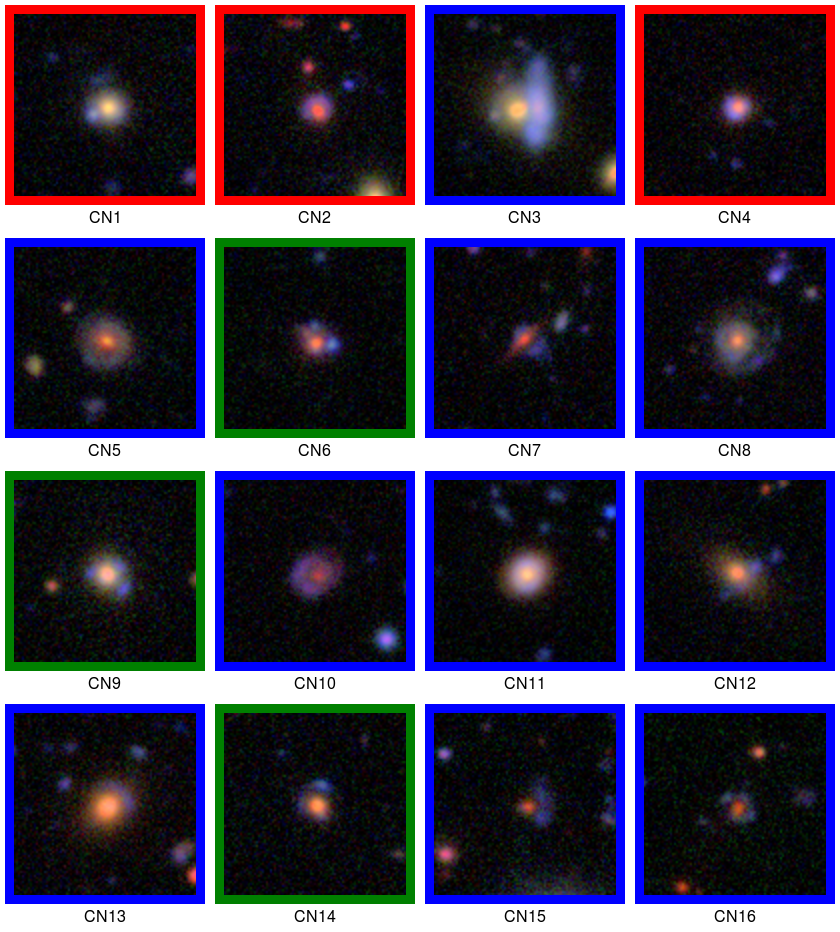}
\caption{New potential galaxy-galaxy lenses detected by the
ConvNet-based lens finder. \(gri\) composite images are shown, with the
border indicating the method of discovery: red = all-survey search, blue
= catalogue based, green = both. The sources are described in
Table~\ref{tbl:new_lenses}.}\label{fig:new_candidates}

\end{figure*}

The sample contains 117 of the 565 previously known lenses, including
29/103 confirmed lenses. After visual inspection of the 2,465 candidates
we identify a further 266 candidates as possible or probable lenses,
including 249 possible, 15 probable and four definite lenses. The
candidates with scores \(\ge 2\) are presented in
Figure~\ref{fig:new_candidates} and Table~\ref{tbl:new_lenses}. 12 of
the galaxies have photometric redshifts, ranging from 0.42 (CN1) to 1.13
(CN16).

Figure~\ref{fig:foundmissed} depicts a subset of test set lenses and
confirmed lenses recovered in the candidate set, and of those not
recovered. There are several example of highly elliptical deflectors
(potentially edge-on spirals) and redder sources amongst the missed
lenses, which differ from those generated for the training set. There
are also several cluster-scale lenses, which were also not included in
our simulations.
In colour-magnitude space, there is not clear difference between the lenses
found and those missed. However, the robot performed slightly better 
on lenses rated as higher quality by astronomers. Of the 59 \textsc{SpaceWarps}
test set lenses, 30 were of grade less than 2, and 29 were of grade 2 and above. 
Of the 2+ lenses, our system recovered 14/29 (46\%), and of those <2, 9/30 (30\%).  

\section{Discussion}\label{sec:discussion}

\subsection{Training networks on simulated
images}\label{training-networks-on-simulated-images}

From the results presented we conclude that convolutional neural
networks trained on synthetic lenses are able to develop detectors for
lensing features and identify real lenses in survey image data. Our
two training sets were designed to simulate strong lenses with
parameters such as brightness and Einstein radius similar to those
lenses present in CFHTLS and discovered by the human volunteers of the
\textsc{SpaceWarps} project. 

As described in Section~\ref{sec:test_set}, we also used a test set 
consisting of the 59 \textsc{SpaceWarps} lenses
to aid in setting some of the search parameters, namely ConvNet
score thresholds.  As the ConvNets are,
by design, optimized to detect objects similar to the training set, our
robot will not in general succeed at finding lenses that are, for
instance, significantly larger in the sky, are of different colour, have
point instead of extended sources (i.e.~QSOs) or are not galaxy-scale
lenses (i.e.~clusters).\footnote{We have created training sets that
  include cluster-scale lenses and lensed QSOs, with preliminary results
  that indicate that these different morphologies are also learnable,
  but do not present that work here.}

Unlike a human, the ConvNets demonstrate little ability to generalise
the features they have learned to situations that differ from those
presented in the training set on which they were optimised. A human
being might spot a lens that has the same shape as other examples but
differs in colour (for instance, a red-red lens) and decide it is an
unusual instance of the class; the ConvNet probably will not, as the
convolutional kernels it has developed weight the contributions of the
colours according to their prevalence and significance in the training
set. Although there is ongoing research into classes of machine learning
techniques where this weakness is addressed - for instance, Bayesian
Program Learning \citep{lake_human-level_2015}, which models visual
concepts as simple probabilistic programs - this is an intrinsic feature
of ANNs.

A sufficiently complex ANN can be expected to learn more than just the
general morphology of a gravitational lens. Any simplifications or
unphysicalities in the simulations are likely to be learned by the
network. Where an ANN is trained properly, the network's weights will
efficiently encode the significant features of the training set. In the
case of real-world images, finding such an encoding is a highly complex
task, though one at which ANNs have proved highly successful. In the
case of simulations, the most efficient such encoding would be the
parameters of the simulation code itself; if an ANN is trained on a
sufficiently large number of simulated images, it can be expected to
become proficient at determining whether an example image could be
generated by a given simulator. This type of overfitting to the training
set is a hazard of using simulated training data, and so good classifier
performance on simulations should be treated with caution. Our own 
results, where excellent performance on simulated lenses was not 
replicated on real data, underscores this point.

In this work we deployed ConvNets with relatively simple architectures.
Training much deeper networks is feasible. The motivation for deeper,
more complex networks is the extraction of a richer feature set from the
training data, at the expense of training time, a larger training set,
and potential difficulties getting training to converge. Because of the
high accuracy on simulated training, validation and test sets, we did
not explore deeper networks which we deemed likely to merely overfit
further to the simulations. Future work will need to explore whether
deeper networks have an impact on the performance on genuine
astronomical image data, especially as the simulations improve and
become more complex.

\subsection{Using an ensemble of Neural
Networks}\label{sec:using_ensemble}

The use of ensembles of neural networks in classification or regression
problems is a common practice in machine learning
\citep{hansen_neural_1990, zhou_ensembling_2002}. According to the
standard procedure, multiple networks are trained using different
subsets of the training set and then (for example) averaged together to
produce a final score for a given input. In complex computer vision
applications this typically provides a small edge over any single
network. In the case of the lensfinder, this approach is of little value
as the network is already able to classify simulated training,
validation and test sets with very high accuracy. The networks'
performance drops when generalised to real data.

We used simulated lenses from two different sources to train our
networks. Each simulation code took an approach that was physical and
realistic but the resulting outputs were slightly different in
appearance. By combining the two, we smooth out some of the less
realistic features of each simulated set. We find that in the case of
our networks, a union of two candidate sets provided good results; we
only look at candidates that triggered both networks, ignoring those
that were flagged in one network only as likely false positives. Other
functions, such as a mean score or max of the two networks' scores, are
also feasible approaches. Although a detailed investigation of the
networks' internal activations is complex and beyond the scope of the
current work, it appears that the ensemble process is effective as the
networks tend to agree most often on a high score where stark lensing
features are apparent, but false positives are triggered in different
ways.

This approach of constructing multiple simulations and training networks
separately may have application in other areas of astronomy where
examples are few and simulations are required to build a training set.

We note that combining the two different sets of simulations into one
and retraining the network on the combined set did not yield
satisfactory results. The combined training set, with an accuracy in
training of \textasciitilde{}70\%, was strictly inferior to either of
the separate training sets. There is no theoretical barrier to achieving
similar accuracy as the ensemble with a single ConvNet of sufficient
complexity, since the logic we employed of developing two sets of
feature maps and weighting the outputs could be replicated within the
connections of an ANN. Further experimentation with the training sets
and the use of a network with additional layers and more artificial
neurons is required.

\subsection{Catalogue-based versus all-survey
search}\label{catalogue-based-versus-all-survey-search}

Searching the entire survey - that is, feeding every pixel of the
imaging data through the networks, regardless of the presence of a
likely lensing galaxy - has the advantage that the maximum completeness
is attainable, since no assumptions are made in selecting a subset of
sources to examine. Unusual lenses of high scientific value (complex
morphologies, multiple lensed sources, dark matter-dominated deflector)
are more likely to be excluded by a catalogue selection, and inaccurate
catalogue photometry may exclude others, such as the case where a bright
blue source substantially shifts the colour of the foreground galaxy
within the photometric aperture used by the catalogue source extractor.
The volunteers for the \textsc{SpaceWarps} CFHTLS blind search were not restricted to postage
stamps of ETGs but shown larger fields containing hundreds of sources,
so they were freely able to nominate candidates that differed from the
typical morphologies and colours used in training examples.

The disadvantages of such a wide search with automated algorithms are 
several. Firstly, it is
computationally expensive, requiring months of GPU time for a large
survey area. Secondly, lower purity in a candidate set can be expected
since a much larger number of candidate images are examined (at least
two orders of magnitude) per genuine lens, accumulating more false
positives. Finally, unless the training set explicitly includes unusual
examples (such as a lens with no visible deflector) they are unlikely to
be detected in any case by the networks.

Restricting the search for strong lenses to preselected potential
deflector galaxies simplifies the search in several ways. Firstly, it
reduces the search area enormously, and presents a much purer (though
inevitably less complete) set of images to the network. Secondly, a
catalogue-based search simplifies the generation of a training set. Both
the positive and negative training images can all be centered on a
bright elliptical galaxy, since all future candidate images will follow
the same convention.

The catalogue-based search was significantly more efficient in recovering
known lenses by producing candidate sets of considerably higher purity
(precision) for a particular completeness (recall) value as measured
against known lenses in the sample. Although the networks trained for
the whole-survey search had a lower false positive rate of 1 in 35,000
versus 1 in 671 for the catalogue-based search, this can be explained by
the relative ease of distinguishing galaxy-galaxy lenses from empty sky,
spiral galaxies, stars and other unlikely candidates.

Of the 63(14) known(confirmed) lens candidates detected in the
all-survey search, 44(11) of them were included in the catalogue sample.
Of the six novel candidates we rate as grade \(\ge 2\) in the all-survey
search, three of them were also detected in the catalogue-based search.

\subsection{Quantifying the performance of the
robot}\label{quantifying-the-performance-of-the-robot}

The outputs of the ConvNets are real numbers \(\in (0,1)\) which we
interpret as a measure of confidence or of candidate quality. The
minimum confidence level for which we include candidates in our output
set is a free parameter in this range. Depending on the values chosen,
the ConvNets are able to produce a set of candidate images of almost
arbitrary size but with highly variable purity and completeness. In
practical terms this means that, depending on the particular
application, the robot can produce larger or smaller candidate sets by
making trade-offs in the purity and completeness.

Analysis of the performance of a lens-finding robot is complicated by
several factors. Performance is optimal when it finds all potential
lenses (lens grade \textgreater{} 0) and no false positives (grade =
0). Performance degrades both as the number of false positives increases
and the fraction of true positives recovered decreases. We can fully
populate the confusion matrix, as defined in Table~\ref{tbl:confusion},
for our system
as applied to training, validation and test sets of simulated lenses.
Similarly, we can measure performance against the test set of
\textsc{SpaceWarps} candidates, since the lens candidates are all known,
as presented in Table~\ref{tbl:confusion_testset}.

It is the robot's performance on real imaging data that is of most interest.
On the wider search, quantifying the performance is more complicated, as
candidates which are not previously reported as lenses in other searches
may not be considered false positives if they are sufficiently
interesting. Similarly, we do not know how many lenses remain undetected
in the survey and so the false negative rate is likewise uncertain. With
this caveat in mind, we use the data from the test set, combined with
more approximate metrics gained from the wider search in our
consideration of our robot's performance on the CFHTLS sky.

\hypertarget{tbl:confusion}{}
\begin{table*}
    \parbox[t][][t]{.45\linewidth}{\centering
\begin{tabular}{@{}lll@{}}
\toprule
\ & Condition Positive (CP) & Condition Negative (CN) \\\midrule
Test positive & True Positives (TP) & False Positives (FP) \\
Test negative & False Negatives (FN) & True Negatives (TN) \\
\bottomrule
\end{tabular}
\caption{\label{tbl:confusion}Template for the confusion matrix of a classification algorithm.}
}
\hfill
    \parbox[t][][t]{.45\linewidth}{\centering
\begin{tabular}{@{}lll@{}}
\toprule
\ & Actual positive & Actual negative \\\midrule
Test positive & 25 & 3 \\
Test negative & 34 & 153,459 \\
\bottomrule
\end{tabular}
\caption{\label{tbl:confusion_testset}Confusion matrix for the all-survey test set containing 153,000 images
including 59 \textsc{SpaceWarps} lens candidates. The candidate set is produced by
taking all candidates where the two ConvNets give a score \(s > 0.95\).}
}
\end{table*}

By varying the free parameters \(t_1\) and \(t_2\)
(Equation~\ref{eq:hyper}) we can more fully explore the balance between
precision and recall afforded by our robot. For any point in the
(\(t_1, t_2\)) space we can plot the rate of true positives (recall) and
false positives (False Positives/Condition Negatives) to produce a
ROC curve as presented in Figure~\ref{fig:ROC_training} and
Figure~\ref{fig:response}. The ideal curve includes the point
(\(x = 0, y = 1\)) where all positive examples are detected with no
false positives, and has an area under the curve (AUC) of 1. The worst case 
case scenario, where the robot guesses randomly, lies on the line \(y = x\).

To construct a confusion matrix and estimate purity and completeness
(precision and recall) on the real survey imaging data, we use the 565
previously reported lens candidates as an approximation of the complete
sample of findable lenses in the survey.

To estimate purity, we also consider interesting novel candidates. Any
candidate to which we assign a grade \textgreater{} 0 after visual
inspection we consider a true positive; other candidates we count as
false positives. Of our final catalogue-based candidate set of 2465 lenses, 117 had been
previously identified as potential lenses, including 23/57 (40\%)
\textsc{SpaceWarps} lenses and 29/103 (28\%) confirmed lenses. The
authors agreed that 364 candidates were of grade 1 or greater, including
98 of the 117 reported lenses in the sample; 526 images were graded with
score \textgreater{} 0 by at least one reviewer, including 102
previously known lenses, meaning that between 379 and 537 of the sample
were interesting candidates. Thus when applied to CFHTLS data, our robot
produced a sample with a purity of between 15-22\%. The 117 of 565 known
lenses (22\%) returned in our sample, including 28\% of confirmed
lenses, gives us an estimate of completeness with respect to the highest
quality detectable lenses.

Our sample included at most 2086 false positives from a set of 1.4
million true negatives tested, for a false positive rate of 0.15\%, or
one in 671.

Although we included all the known lenses and candidates previously
discovered in CFHTLS image searches in our analysis above, a significant
number of them are cluster-scale lenses, have non-elliptical deflectors
or red sources, or are extremely faint according to a subjective ranking
by the authors. These differ significantly from the examples simulated
in the training sets we generated. We estimate approximately 15\% of the
sample, both confirmed and unconfirmed, meet one or more of these
conditions. Excluding these from the known lenses, our lens finder
recovered 25\% of previously reported and 33\% of confirmed lenses; this
may better approximate the performance of a robot engineered
specifically for lenses of a particular morphological class,
i.e.~galaxy-galaxy lenses, and give some indication of gains that might
be realised with a training set that includes greater morphological
diversity.

\subsection{Context and practice}\label{context-and-practice}

The ConvNets were more prone to identify clear false positives than a
professional astronomer or even a human volunteer with minimal training;
for instance, across the whole survey each ConvNet identified of order
\(~10^7\) candidates with confidence \(> 0.5\), amounting to a few
percent of images tested. A key challenge was how to best use the
classification data to minimise false positives, i.e.~candidates we
judge very unlikely to be lensing systems. Intrinsic to the lens-finding
problem is the rarity of lenses on the sky. If we expect approximately
one lens per square degree of sky \citep[as per][]{treu_strong_2010},
then in an all-survey search of the scale we conducted, a false positive
rate of only 1 in 1000 would mean that quality lens candidates would be
outnumbered by false positives at a ratio of about 4000:1.

\begin{figure*}
\centering
\includegraphics[width=0.95000\textwidth]{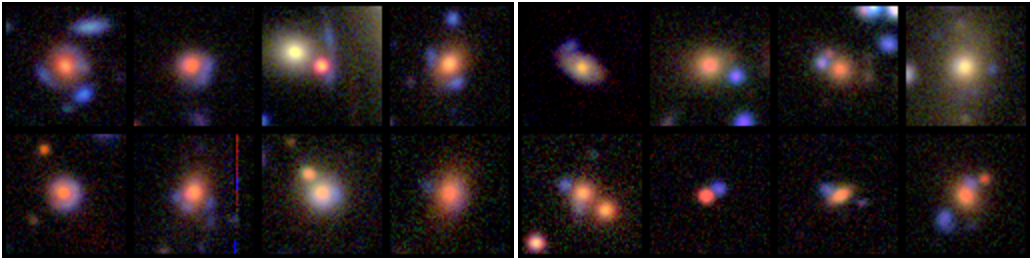}
\caption{An example of a candidate set with high purity (50\%) but low
completeness (1.4\%). By restricting the search to sources
with \(19.5 < \textrm{mag}_i < 20.5\) and \(1.8 < g - i < 2.5\),
and setting the ConvNet thresholds such that only a small candidate set
is returned, we produced this set of 16 candidates including eight 
previously known lenses (left).}\label{fig:pure}

\end{figure*}

An alternate way to examine the trade-off between precision and recall
is presented in Figure~\ref{fig:response}, which plots the fraction of
lenses recovered in the candidate set against the average size of the
candidate set per square degree of sky. Assuming an astronomer could
search these candidates for quality lenses at a rate of 60 per
minute\footnote{When presented as 10x10 montages containing 0 to a few
quality candidates, we found this rate to be feasible.}, the amount of
time required to process the whole survey's candidates is indicated on
the right. For a whole-survey search, to recover 50\% of lenses one
would need to examine approximately 250 candidates per square degree,
0.2\% of the field's area. To recover a minimum of one high-quality
candidate, examining a candidate set of a few hundred postage stamps for
the whole survey would be required. The catalogue-based search is more
efficient, with 50\% recall achievable with only 40 candidates per
square degree. By restricting candidate sources even further,
to \(19.5 < \textrm{mag}_i < 20.5\) and \(1.8 < g - i < 2.5\),
and setting the thresholds aggressively we were able to generate small, 
pure but highly incomplete candidate sets, for instance
a set containing only 7 candidates for the entire survey but including 4 known lenses 
(a purity of 57\%). The lensfinder can be used to quickly find some of the
best quality lenses, but astronomer time required grows quickly as desired completeness 
increases (see Section~\ref{sec:catalog_based} and
Figure~\ref{fig:pure}).

Table~\ref{tbl:time_estimates} presents estimated astronomer time for recovering 
galaxy-galaxy lenses from DES and LSST assuming a candidate finder with
similar purity and completeness characteristics as the one we applied to 
CFHTLS. \citet{collett_population_2015} estimates 1,300 detectable lenses in DES
and 62,000 in LSST. Recovering the majority of them from a survey of LSST's size
could require months of astronomer time categorising candidates. Although 
the efficiency of the lensfinder we developed can no doubt be significantly improved
upon, these figures highlight the need (and opportunity for) new algorithms 
that can more efficiently extract the most interesting candidates 
from the surveys' imaging.

\hypertarget{tbl:time_estimates}{}
\begin{table}
\centering

\caption{\label{tbl:time_estimates}
Estimates of astronomer time required to find 20, 40, 60 and 80\% of detectable lenses in the Dark Energy Survey (DES) and Large Synoptic Survey Telescope (LSST) surveys using the ConvNet method assuming similar performance to that presented in this paper. Estimates from Collett (2015) suggest 1300 findable galaxy-galaxy lenses in DES and 62000 in LSST.}
    
\begin{tabular}{@{}lllll@{}}
\toprule

Survey & 20\% & 40\% & 60\% & 80\% \\
\midrule
CFHTLS & 3 min & 30 min & 5 hours & 4 days \\
DES & 30 min & 5 hours & 2 days & 44 days \\
LSST & 4 hours & 3 days & 22 days & 1.5 years \\

\bottomrule
\end{tabular}

\end{table}

\section{Conclusions}\label{sec:conclusions}

We present an application of convolutional neural networks to the
automated identification of potential strong lenses. Deep learning
techniques have advanced computer vision in many other fields and have
already found many promising applications in astronomy. Our work
demonstrates that a ConvNet can extract morphological and colour-space
features from examples of strong lenses, and also that the use of
simulations in training sets can be used to train a lens-finding robot
to be of practical use on real astronomical data. For a survey covering
171 degrees to an r-band depth of \textasciitilde{}24.83 we were able to
generate a candidate set that was 28\% complete in terms of confirmed,
findable lenses; 15\% pure with respect to possible lenses; and at 
\textasciitilde{}14 candidates per square degree, could be
visually inspected by a human in under an hour. 
The ConvNet-based method allows for higher completeness at 
the cost of decreasing purity.

Using the pipeline developed here we find that a ConvNet-based
lensfinder, under the conditions of the CFHTLS and using a catalogue-based
search with optimised settings, could produce a candidate set that
yielded of order one good quality lens candidate per square degree of
survey sky with an investment of astronomer inspection time of under
half a minute each. We are optimistic that this rate of discovery can be
replicated or exceeded in other extant and future image surveys. 

The rate of discovery achieved with the ConvNets compares favourably 
to previous robotic searches. Our lensfinder proved more efficient 
than \citeauthor{more_cfhtlsstrong_2012}'s \citeyearpar{more_cfhtlsstrong_2012} 
CFHTLS lens search using \textsc{arcfinder}, 
which required visual inspection of \textasciitilde{}150,000 
candidates for a final sample of 127 quality candidates, and is on par with 
\textsc{RingFinder} \citep{gavazzi_ringfinder:_2014}, where
the algorithm returned 0.4\% of sources examined, for a sample size of 
2500 candidates. Gavazzi et al estimate 40\% completeness based on
simulations and follow up of a sub-sample of candidates;  
our catalogue-based search returned 0.2\% of ETGs examined and we place
a lower bound on completeness of 28\%. The analysis presented in 
Section~\ref{quantifying-the-performance-of-the-robot} suggests that
at current performance, a completeness of 40\% of findable
lenses is achievable with a candidate set of \textasciitilde{}8000 candidates,
requiring about three hours per astronomer to sift. We expect that
with more realistic training sets and different ensemble strategies
this performance can be significantly improved upon.

CFHTLS was chosen as a yardstick to evaluate ConvNet performance as it
has been extensively searched for lenses using other methodologies.
Given the performance we obtained, this method would appear to be a
promising way to explore new fields. Using synthetic training sets
generated with parameters matching other current and upcoming surveys
such as DES and LSST, the convents can be retrained and readily applied
to another data pipeline. As well as generating candidate sets from
these surveys, this method could also be used to produce candidates for
revision by citizen scientists, complementing efforts like
\textsc{SpaceWarps} by screening candidates but relying on human
volunteers and the \textsc{SpaceWarps} quality algorithm to purify the
sample before review by astronomers.

Given the demonstrated effectiveness of convolutional neural networks in
computer vision applications such as classification and detection of
everyday objects, it has proven to be a safe assumption that, if there
is enough information in an image for a human expert to extract meaning,
a properly-designed ConvNet is likely to also converge on a suitable
feature extraction strategy. The \textsc{SpaceWarps} project, by
providing rapid training to human volunteers, demonstrated that enough
information is present in images of galaxies under the conditions of
CFHTLS resolution and seeing.

The need for simulated lenses to train Convolutional Neural Networks
is potentially a limiting factor in their utility in practice, as
they can only ever be as good as the simulations are realistic.
Our results underscore the risk that Convolutional Neural Networks 
trained on simulations may learn to overfit to the peculiarities of 
the simulation code, and that good performance on a simulated 
training and test set will not, as a rule, translate directly to
the real universe. Reports of good performance measured in simulations 
only should therefore be treated with caution, as it may merely indicate 
that the network has learned to spot a subtle bias or artifact 
present in simulated lenses rather than physical lensing features.

Futher work includes refining the training sets to reduce false
positives and to better detect lenses with a wider range of
morphologies, including lensed QSOs and cluster-scale lenses; testing an
increase in the size of the training sets and the number of
convolutional layers in the network; and retraining the networks with
good candidates and false positives identified in visual inspection.
Exploring the ensemble approach further, for instance with training sets
composed of simulated lenses in discrete bins of source magnification
and Einstein radius, may aid in developing a detector capable of
discriminating the brightest and most promising lenses more efficiently.

Convolutional neural networks have also recently been employed for lens
finding in the Kilo-Degree Survey (KiDS) \citep{petrillo_finding_2017} and simulations of LSST
\citep{lanusse_cmu_2017}. The differences in survey data quality and the
types of lens targeted make a direct comparison difficult. However, our
method of using multiple training sets and ConvNets appears to give more
probable lens candidates per square degree than the Petrillo et al
ConvNets, although at the cost of a factor four more human
classification time. This lower purity is unlikely to be a relative
deficiency of our method; we targeted smaller Einstein radius lenses
where seeing makes the classification problem much harder.


\bibliographystyle{mnras}
\renewcommand\refname{References}
\bibliography{./cfhtls_convnet.bib}

\label{lastpage}
\end{document}